# A novel mechanism for energy activation in biomolecules

Shanshan Wu and Ao Ma*

Department of Bioengineering

The University of Illinois at Chicago

851 South Morgan Street

Chicago, IL 60607

*correspondence should be addressed to:

Ao Ma

Email: aoma@uic.edu

Tel: (312) 996-7225

**Abstract**

An activated process consists of energy activation and barrier crossing; the former is a prerequisite for the latter. Barrier crossing has been studied extensively, but energy activation has been overlooked due to a lack of means to gauge its progress. We define reaction stability $p_S$ as the probability that reactive trajectories pass a vicinity in phase space; it enabled us to analyze energy activation of a biomolecular isomerization. This process follows a mechanism fundamentally different from presumed mechanisms in standard reaction rate theories—it features accumulation of high kinetic energy in reaction coordinates, achieved by precise synergy between them coordinated by momentum space.

Activated process is essential to all molecular systems, ranging from chemical reactions of small molecules to conformational dynamics and enzymatic catalysis of proteins. For proteins, all the functionally important processes are activated processes because it gives them well-defined rates, which is critical for the functional roles of proteins in the cellular context because proper timing is key for functioning. The defining feature of an activated process is that the system needs to cross an activation barrier that is much higher than the thermal energy $k_B T$, where $k_B$ is the Boltzmann constant and $T$ is temperature. There are two critical questions concerning the mechanism of an activated process: (1) how the reaction coordinates acquire sufficient energy to cross the activation barrier, and (2) how the reaction coordinates cross the barrier once they acquired adequate energy. The first concerns energy activation and the second concerns barrier crossing [1]. The former is a prerequisite for and must precede the latter.

Barrier crossing has been intensively studied both theoretically, through efforts in refining the reaction rate theories [1-3], and numerically, with molecular dynamics simulations. The complexity of numerical studies has grown with the increase in computational power: from early studies of gas phase reactions [4], to later studies of reactions in solution [5], to biomolecular dynamics in the past two decades with the help of transition path sampling method [6, 7] to harvest unbiased reactive trajectories and committor [8] to parameterize the progress of barrier crossing.

In contrast, energy activation only received attention at the early stage of the development of reaction rate theories, with three lines of ideas. In transition state theory (**TST**), the problem of energy activation is bypassed by the quasi-equilibrium assumption between transition state and reactant. In Lindman's mechanism for unimolecular reactions in gas phase, it was assumed that collisions between a reactant molecule and buffer gas provide the reactant with extra energy, which is quickly equilibrated among all the coordinates via fast intra-molecular vibrational redistribution [1]. The subsequent barrier crossing of the high-energy reactant is under this condition of equipartition of energy over the entire molecule. In Kramers theory, it was assumed that the reaction coordinate acquires energy from thermal fluctuations of the solvents, modeled as a random force [9]. Therefore, the reaction coordinate can climb up the activation barrier during rare fluctuations in which the random force increases the total energy of the reaction coordinate.

The overwhelming interest in activated processes in proteins over the past few decades has presented new challenges as well as opportunities. Since protein reactions always occur in solutions, TST and Kramers theory were widely adopted. However, a complex molecule like a protein differs fundamentally from a simple molecule, on which Lindman's and Kramers' ideas were based. The large size of a complex molecule means it has sufficient degrees of freedom that it can provide the reaction coordinates with adequate energy for activation on its own. In contrast, simple molecules require an external energy source, be it buffer gas or solvents, for activation. More importantly, in complex molecules reaction coordinates and the thermal bath are connected by chemical bonds and the interactions between them involve a wide range of spatial scales. Bonded interactions impose complex and strong constraints on the motions of bath coordinates. This is in stark contrast with buffer gas molecules that essentially undergo free motions and solvent molecules that undergo a combination of libration and diffusion. The existence of interactions with different spatial scales presents a much more complex situation than the linear coupling between reaction coordinates and thermal bath assumed in standard reaction rate theories. These unique features of complex molecules make it possible that their energy activation may follow a mechanism fundamentally different from mechanisms suggested by Lindman's and Kramers' ideas. This could be the reason behind the challenges encountered in applications of TST and Kramers theory to activated processes in proteins [10, 11].

One instance that hints at this issue is an unexpected observation from our previous study of a prototype of biomolecular isomerization dynamics—the $C_{7eq} \to C_{7ax}$ transition of an alanine dipeptide in vacuum [12]. Alanine dipeptide is the smallest molecule in which the non-reaction coordinates constitute a large enough thermal bath for activation, namely, it is the smallest example of complex molecules. Previous studies have identified dihedrals $\phi$ and $\theta_1$ (Fig. 1) as the essential reaction coordinates for this process—they are sufficient for determining the value of committor $p_B$ [7, 13]. Committor is the probability that a dynamic trajectory initiated from a configuration space point, with momenta drawn from Boltzmann distribution, to be a reactive trajectory. It is the reaction probability in configuration space and provides a rigorous parameterization of the progress of barrier crossing. Our previous study showed that the activation barrier is located on the path of $\phi$, and $\theta_1$ helps $\phi$ to cross this barrier by directly transferring kinetic energy to $\phi$ [12, 14].

The unexpected observation concerns failed attempts of $\phi$ to cross the activation barrier. As shown in Fig. 2, $\phi$ already reached the critical value that marks the onset of a successful transition at t = 1.03 ps. However, instead of moving forward and crossing the activation barrier, $\phi$ reversed its direction and receded back to the reactant basin. A closer examination revealed that the critical difference between the failed and the succeeded barrier crossing is the position of $\theta_1$. The incorrect position of $\theta_1$ makes the force from $\theta_1$ to $\phi$ (Fig. 2), defined as $\Delta F_{\theta_1 \to \phi} = -\int_{t_0}^{t_1} \frac{\partial^2 U(\vec{R})}{\partial \phi \partial \theta_1} d\theta_1$ ($U(\vec{R})$ is the system potential energy), strongly repulsive, which drives $\phi$ away from the activation barrier and reverses its direction of motion.. Even though the force that pushes $\phi$ up the activation barrier is orders of magnitude stronger than what $\phi$ normally experiences during a successful transition, $\Delta F_{\theta_1 \to \phi}$ increases much faster until it exceeds the force that facilitates $\phi$'s barrier crossing and reverses the direction of the total force acting on $\phi$. These results show that the motions of $\phi$ and $\theta_1$ need to be precisely coordinated: it appears that $\theta_1$ acts as a gating mechanism on the motion of $\phi$.

The observation above suggests that some critical events occurred in the region where $p_B = 0$ ubiquitously, the region that $p_B$ cannot parameterize. This indicates a region where the momentum space is critical for activation [15], as $p_B$ provides rigorous parameterization of activation in configuration space. To understand the difference between a failed attempt for barrier crossing and a successful one, we need a rigorous parameterization of this region. The first step is to find a proper reference point. The natural choice is a phase space point $\Gamma_0$ on an existing reactive trajectory. If we perturb the momenta of $\Gamma_0$ slightly, we obtain a phase space point $\Gamma_0'$ in its close vicinity. The probability that a dynamic trajectory launched from $\Gamma_0'$ is a reactive trajectory is less than 1. This probability reflects the likelihood that reactive trajectories pass the vicinity around $\Gamma_0$.

Based on this idea, we can define a new parameter $p_S$, which we call the reaction stability. For a given phase space point $\Gamma_0$ on an existing reactive trajectory, we can generate an ensemble of phase space points $E(\Gamma_0)$ in its vicinity. A point $\Gamma_0'$ in $E(\Gamma_0)$ is generated from a small random perturbation $\epsilon$ (e.g. 20%) to the momenta of $\Gamma_0$. From $\Gamma_0'$ we can launch a dynamic trajectory and check if it is reactive. In this way, we obtain the probability of trajectories launched from $E(\Gamma_0)$ to be reactive, which is the value of $p_S$ for $\Gamma_0$. A large value of $p_S$ suggests that $\Gamma_0$ lies in a phase space region that has dense populations of reactive trajectories, so that a perturbation to $\Gamma_0$ will more likely land on a phase space point on another reactive trajectory. Along a given reactive trajectory, the system in general moves from region with low density towards region with high density of reactive trajectories. Therefore, $p_S$ provides a rigorous parameterization of the progress of activation in phase space. The region with $p_S \in (0,1]$ and $p_B = 0$ precedes the region with $p_B \in (0,1]$; the former corresponds to the energy activation phase and the latter corresponds to the barrier crossing phase.

To analyze the mechanism for energy activation, we use the energy flow theory we recently developed. Within this theory, we define both potential (**PEFs**) and kinetic (**KEFs**) energy flows during an activated process. The PEF through a coordinate $q_i$ is its work:

$$\Delta W_i(t_1, t_2) = -\int_{q_i(t_1)}^{q_i(t_2)} \frac{\partial U(\vec{q})}{\partial q_i} dq_i \quad (1).$$

According to Eq. (1), $\Delta W_i(t_1, t_2)$ is the change in the potential energy of the system due to the motion of $q_i$ along a dynamic trajectory in the time interval $[t_1, t_2]$. It is a projection of the change in the total potential energy onto the motion of $q_i$. Therefore, it is a measure of the cost of the motion of $q_i$ in terms of potential energy. Accordingly, the change in the total potential energy of the system can be decomposed into PEFs through different coordinates: $\Delta U(t_1, t_2) = U(t_2) - U(t_1) = -\sum_{i=1}^{N} \Delta W_i(t_1, t_2)$, where the summation is over all coordinates of the system. A major finding from our previous PEF analysis was that reaction coordinates are the coordinates with high PEFs during barrier crossing [12].

The KEF through a coordinate $q_i$ is [14]:

$$\partial_v K_i = \frac{\partial K}{\partial \dot{q}_i} d\dot{q}_i + \left(\frac{\partial K}{\partial q_i}\right)_{\vec{q}',\vec{v}} dq_i = dt\left[p_i \ddot{q}_i + \left(\frac{\partial K}{\partial q_i}\right)_{\vec{q}',\vec{v}} \dot{q}_i\right] \quad (3),$$

where $K$ is the system kinetic energy, $\vec{q}' = (q_1, q_2, \cdots, q_{i-1}, q_{i+1}, \cdots, q_N)$ is the system position vector in internal coordinates with $q_i$ removed, and $\vec{v} = (\dot{q}_1, \dot{q}_2, \cdots, \dot{q}_N)$ is the velocity vector. Since $\partial_v K_i$ is the change in the system kinetic energy caused by changes in $(q_i, \dot{q}_i)$ alone, which fully describes the motion of $q_i$, it rigorously defines the KEF through $q_i$. Similarly, we have: $dK = \sum_{i=1}^{N} \partial_v K_i$.

To gain mechanistic insights, we need to look at how the PEFs and KEFs of individual coordinates change with the progress of activation. We first project the PEF or KEF onto a projector $\xi(\Gamma)$ that parameterizes the progress of activation, then average over the ensemble of reactive trajectories:

$$\langle \delta A_i(\xi^*) \rangle = \frac{\int d\Gamma \rho(\Gamma) \delta A_i(\xi(\Gamma) \to \xi(\Gamma) + d\xi) \delta(\xi(\Gamma) - \xi^*)}{\int d\Gamma \rho(\Gamma) \delta(\xi(\Gamma) - \xi^*)}$$

$$\langle \Delta A_i(\xi_1 \to \xi_2) \rangle = \int_{\xi_1}^{\xi_2} \langle \delta A_i(\xi) \rangle \quad (4)$$

Here, $\rho(\Gamma) d\Gamma$ is the probability of finding the system in an infinitesimal volume $d\Gamma$ around a point $\Gamma$ in phase space in the transition path ensemble; $\delta(x)$ is the Dirac $\delta$-function; $\delta A_i(\xi(\Gamma) \to \xi(\Gamma) + d\xi)$ is the change in $A_i$ in a differential interval $[\xi(\Gamma), \xi(\Gamma) + d\xi]$; $\langle \Delta A_i(\xi_1 \to \xi_2) \rangle$ is the change in $A_i$ in a finite interval $[\xi_1, \xi_2]$, $\Delta A_i$ can be either $\Delta W_i$ or $\Delta_v K_i$ [12, 14]. For the barrier crossing phase, the optimal projector is $\xi = p_B$; for the energy activation phase, the optimal projector is $\xi = p_S$.

Our results show that the duration of the energy activation phase is generally about 10 times longer than the duration of the barrier crossing phase, suggesting the former is a more complex and important process than the latter. Figure 3 shows the PEFs and KEFs through all the coordinates in the system. As expected, only the two dominant reaction coordinates, $\phi$ and $\theta_1$, experience significant PEFs during energy activation. Moreover, the PEFs through $\theta_1$ during energy activation and barrier crossing are of opposite signs. While $\theta_1$ receives energy through PEFs during barrier crossing, there is a barrier of ~2.5 kJ·mol$^{-1}$ on its path of motion during energy activation. Importantly, the energy flows in $\phi$ cannot start until $\theta_1$ reaches the top of this barrier at $p_S \simeq 0.4$. This explains why $\theta_1$ can act as a gating mechanism on $\phi$: the necessary condition for barrier crossing of $\phi$ is its energy flow, which cannot start before $\theta_1$ crossed its own barrier and became 'ready' to help $\phi$, as shown in ref. [14].

A surprising observation is that kinetic energy plays the dominant role in energy activation for $\phi$. After $\theta_1$ crossed the barrier on its path of motion, kinetic energy starts to accumulate in $\phi$ until it reaches ~10 kJ·mol$^{-1}$, which is about the same as the total amount of potential energy consumed during its barrier crossing. This means that the energy cost for the barrier crossing of $\phi$ is mainly paid by the kinetic energy it gathered during a long energy activation process. In addition, the momentum space is critical in this process, i.e. the proper alignment of momenta of all the coordinates in the system is critical for successfully building up kinetic energy in $\phi$. Moreover, a significant portion (~60%) of kinetic energy accumulated in $\phi$ is due to the loss of kinetic energy in $\theta_1$ and $\psi$. This also explains why the motions of reaction coordinates are highly coordinated as their momenta have to align properly to ensure successful transfer of kinetic energy from $\theta_1$ and $\psi$ to $\phi$.

The mechanism of energy activation discussed above, the persistent accumulation of high amount of kinetic energy in the reaction coordinates, is fundamentally different from the physical picture for energy activation suggested by Lindman's and Kramers theories, which are the foundation of our understanding of activated dynamics. It suggests that energy activation in complex molecules like proteins follows fundamentally new mechanisms. Since alanine dipeptide is the smallest complex molecule, real proteins likely employ energy activation mechanisms that are more sophisticated and effective. This might be the reason behind the extraordinary efficiency of enzymes: proper energy activation is the foundation for protein functions and efficient energy activation leads to efficient functionality.

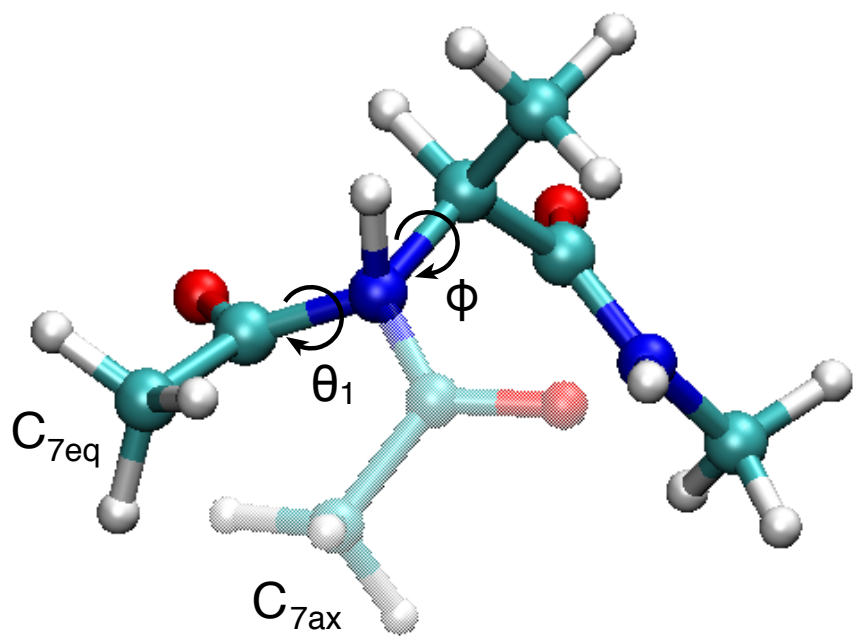

**Figure 1**: Two representative structures of an alanine dipeptide for the $C_{7eq}$ (solid color) and $C_{7ax}$ (semi-transparent color) states. The part of the molecule that does not change in the $C_{7eq} \to C_{7ax}$ transition completely overlaps between the two representations. The essential reaction coordinates $\phi$ and $\theta_1$ are indicated.

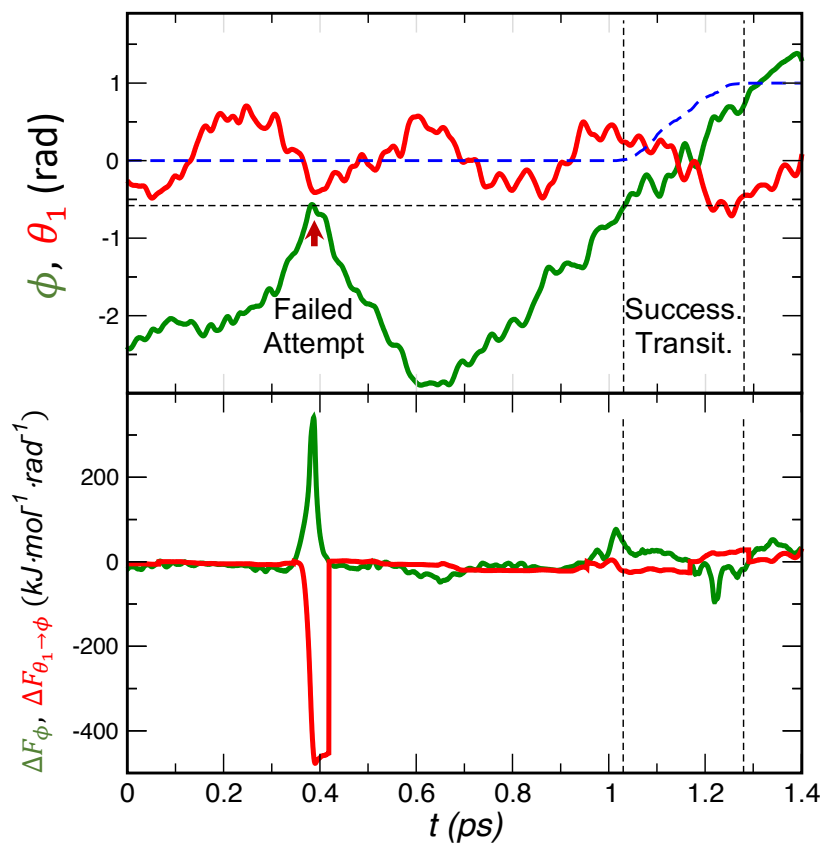

**Figure 2:** (Upper) Time evolution of $\phi$ (green) and $\theta_1$ (red) along a reactive trajectory that includes a failed attempt of barrier crossing marked by an arrow. Blue dashed line: time evolution of $p_B$. The horizontal dashed line marks the critical value of $\phi$ that marks the onset of barrier crossing. The two vertical dashed lines mark the region of the successful barrier crossing. (Lower) Time evolution of the total force acting on $\phi$ ($\Delta F_\phi$; green) and the force from $\theta_1$ to $\phi$ ($\Delta F_{\theta_1 \to \phi}$; red). Note the huge difference in the magnitudes of $\Delta F_\phi$ and $\Delta F_{\theta_1 \to \phi}$ between the failed attempt around t = 0.4 ps and the successful transition during $t \in [1, 1.3]$ ps.

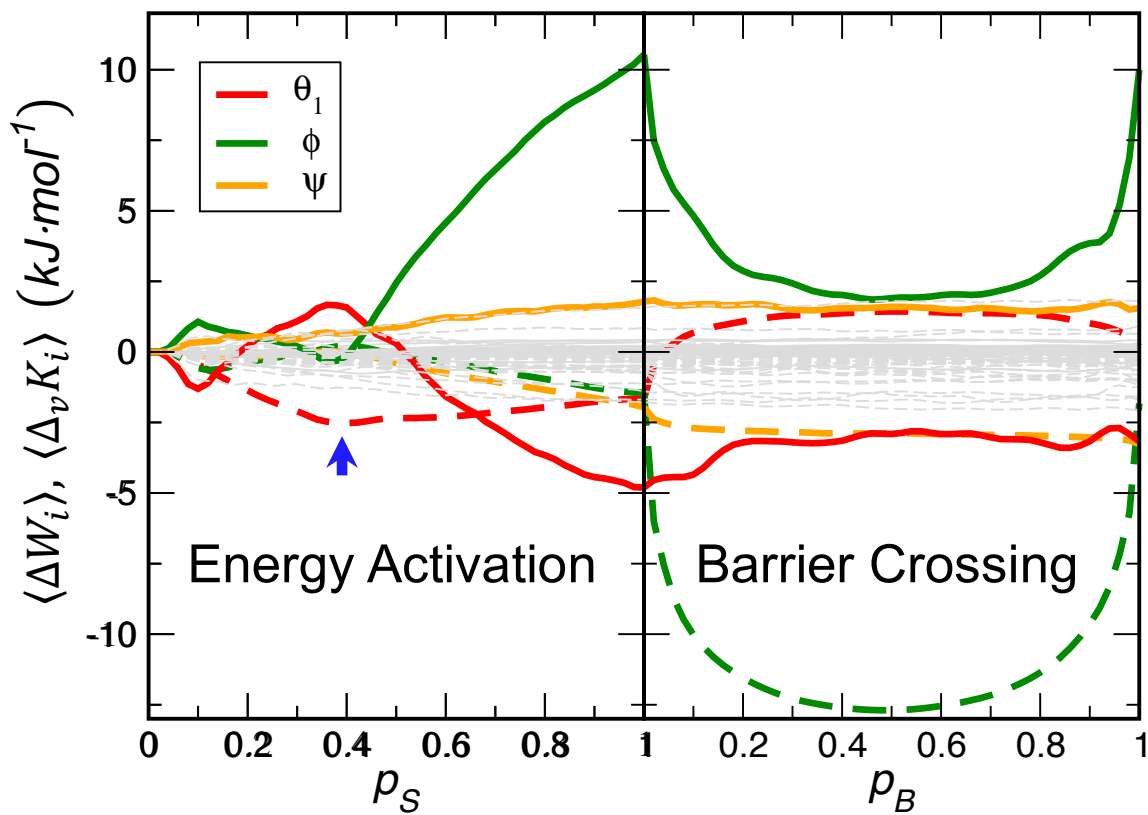

**Figure 3**: The PEFs ($\langle \Delta W_i \rangle$ $i = \phi, \theta_1, \psi$; dashed lines) and KEFs ($\langle \Delta_v K_i \rangle$ $i = \phi, \theta_1, \psi$; solid lines) during energy activation phase (left panel) and barrier crossing phase (right panel). The blue arrow marks the top of the barrier on the path of $\theta_1$ during energy activation. The gray dashed lines are the PEFs and KEFs of the other coordinates (57 in total) in the system.